\documentclass[aps,prl,reprint,letterpaper,amsmath,amssymb, superscriptaddress]{revtex4-1}
\usepackage{graphicx}

\begin{document}
\title{Electron charge and spin delocalization revealed in the optically probed longitudinal and transverse spin dynamics in $n$-GaAs}

\author{V. V. Belykh}
\email[]{vasilii.belykh@tu-dortmund.de}
\affiliation{Experimentelle Physik 2, Technische Universit\"{a}t Dortmund, D-44221 Dortmund, Germany}
\affiliation{P.N. Lebedev Physical Institute of the Russian Academy of Sciences, 119991 Moscow, Russia}
\author{K. V. Kavokin}
\affiliation{Ioffe Institute, Russian Academy of Sciences, 194021 St. Petersburg, Russia}
\affiliation{Spin Optics Laboratory, St. Petersburg State University, 199034 St. Petersburg, Russia}
\author{D. R. Yakovlev}
\affiliation{Experimentelle Physik 2, Technische Universit\"{a}t Dortmund, D-44221 Dortmund, Germany}
\affiliation{Ioffe Institute, Russian Academy of Sciences, 194021 St. Petersburg, Russia}
\author{M. Bayer}
\affiliation{Experimentelle Physik 2, Technische Universit\"{a}t Dortmund, D-44221 Dortmund, Germany}
\affiliation{Ioffe Institute, Russian Academy of Sciences, 194021 St. Petersburg, Russia}

\date{\today}
\begin{abstract}
The evolution of the electron spin dynamics as consequence of carrier delocalization in $n$-type GaAs is investigated by the recently developed extended pump-probe Kerr/Faraday rotation spectroscopy. We find that isolated electrons localized on donors demonstrate a prominent difference between the longitudinal and transverse spin relaxation rates in magnetic field, which is almost absent in the metallic phase. The inhomogeneous transverse dephasing time $T_2^*$ of the spin ensemble strongly increases upon electron delocalization as a result of motional narrowing that can be induced by increasing either the donor concentration or the temperature. An unexpected relation between $T_2^*$ and the longitudinal spin relaxation time $T_1$ is found, namely that their product is about constant, as explained by the magnetic field effect on the spin diffusion. We observe a two-stage longitudinal spin relaxation which suggests the establishment of spin temperature in the system of exchange-coupled donor-bound electrons.
\end{abstract}

\maketitle

The dynamics of localized spins in solids with a rigid crystal lattice is known to be drastically different from that in gaseous and liquid phases. The main features characteristic for solids are the prominent difference between the transverse and longitudinal spin relaxation times in magnetic field and the strong inhomogeneous broadening of magnetic resonance spectra. Both features disappear or are at least strongly reduced for mobile spins, because of the mixing of the motional and spin degrees of freedom and the rapid change of the interacting spin environment.

Semiconductors are ideally suited for changing the spin localization in a controlled way. Electrons localized at shallow impurities can become mobile by increasing the temperature, for example. In a series of semiconductor structures with increasing doping but otherwise identical properties, a transition to metallic conductivity occurs when the impurity concentration exceeds a certain threshold (Mott transition). Due to the spin angular momentum carried by electrons and holes, these modifications allow one to assess the effect of localization on the spin dynamics.

In the present work we study the electron spin dynamics in bulk $n$-type GaAs, which is a prototypical system for optical access to the electron spin states. In particular, the non-equilibrium electron spin lifetime in weak magnetic fields was previously shown to change when the donor concentration $n_\text{D}$ crosses the Mott-type metal-to-insulator transition (MIT) at $n_\text{D}=(1-2) \times 10^{16}$~cm$^{-3}$ \cite{Dzhioev2002}. At low donor concentrations, the spin lifetime is limited by inhomogeneous dephasing in the random nuclear fields. With increasing $n_\text{D}$, the isotropic exchange interaction of electrons causes their coupling with nuclear spins to be less effective, so that the spin lifetime becomes longer, reaching a maximum of $\sim 200$~ns at $n_\text{D}=5 \times 10^{15}$~cm$^{-3}$ (somewhat below the MIT). Above the MIT, the spin lifetime rapidly decreases with increasing electron concentration due to the Dyakonov-Perel relaxation mechanism \cite{Dzhioev2002,Crooker2009,Romer2010,Furis2006,Lonnemann2017}. Indications for a similar trend were reported for the spin dephasing time $T_2^*$ in nonzero transverse magnetic field \cite{Kikkawa1998}. On the other hand, the spin relaxation time $T_1$ in a longitudinal magnetic field falls in the microsecond \cite{Colton2004,Colton2007} or even millisecond \cite{Fu2006,Linpeng2016} range at low donor concentrations, while it is in the submicrosecond range above the MIT \cite{Belykh2016}.

\begin{figure*}
\includegraphics[width=2\columnwidth]{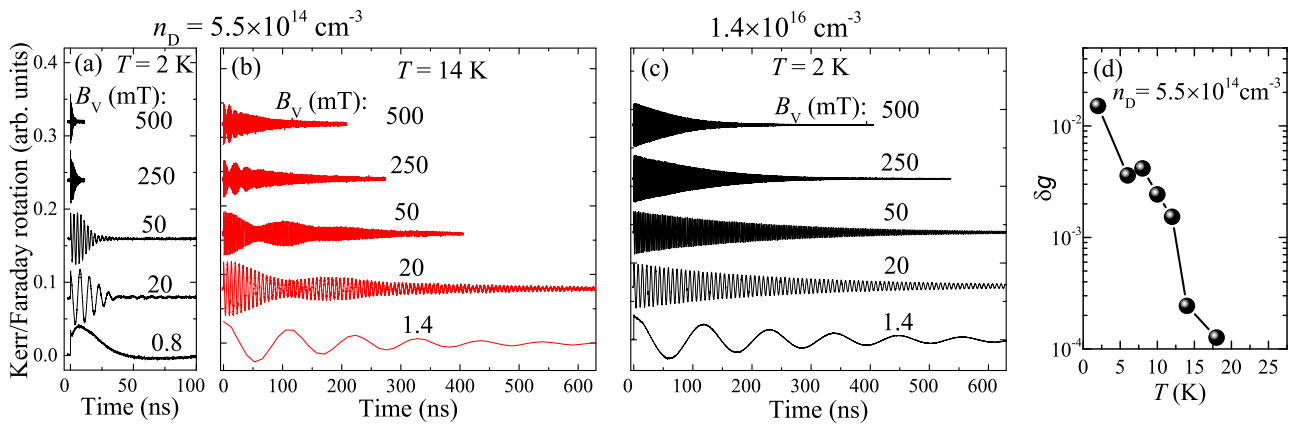}
\caption{(a)-(c) Dynamics of Kerr/Faraday rotation signal for different magnetic fields applied in the Voigt geometry. Panels (a) and (b) correspond to sample with $n_\text{D} = 5.5 \times 10^{14}$~cm$^{-3}$ at $T = 2$ and 14~K, respectively. Panel (c) corresponds to $n_\text{D} = 1.4 \times 10^{16}$~cm$^{-3}$ at $T = 2$~K. (d) Temperature dependence of spread of $g$ factors for $n_\text{D} = 5.5 \times 10^{14}$~cm$^{-3}$.}
\label{FigVoigt}
\end{figure*}

However, because of experimental limitations, so far a comprehensive study revealing the changes in both longitudinal and transverse spin dynamics when crossing the MIT is still lacking. The relevant spin relaxation times range from picoseconds to milliseconds depending on doping concentration, temperature and magnetic field, while standard pump-probe Faraday/Kerr rotation, providing direct access to the spin dynamics, is limited to a few nanoseconds time range. Therefore, measurements based on the Hanle effect (near zero magnetic field) \cite{Dzhioev2002,Dzhioev2002a,Furis2006,Lonnemann2017}, resonant spin amplification \cite{Kikkawa1998} and spin noise (in transverse magnetic field) \cite{Crooker2009,Romer2010} are used to extract long spin lifetimes. These measurements, however, are indirect and do not provide comprehensive insight into complex spin dynamics reflected, e.g, by a nonexponential decay. The longitudinal spin dynamics is usually studied by pump-probe methods based on an analysis of the polarization-resolved photoluminescence, which has a rather low time resolution and is difficult to apply at low magnetic fields \cite{Colton2004,Colton2007,Fu2006,Linpeng2016}. In the present work we overcome these limitations by using the extended pump-probe Faraday/Kerr rotation technique, which allows direct measurement of both the transverse and longitudinal spin dynamics in magnetic fields of any strength with picosecond time resolution over an arbitrary long time range \cite{Belykh2016}.

This technical advancement has led us to qualitatively new findings.
We clearly observe a crossover in the spin dynamics patterns from systems of localized spins to those with delocalized spins when increasing either doping concentration or sample temperature. The crossover is manifested in a strong narrowing of the $g$-factor distribution and dramatic weakening of the magnetic field dependence of $T_1$. Further, we find the unexpected relation $T_1 T_2^*\approx$const which holds even though $T_1$ and $T_2^*$ vary with magnetic field or temperature by up to two orders of magnitude. In the $n_\text{D}$ range just below the MIT we find a double-exponential longitudinal spin dynamics, which reflects the fast onset of internal equilibrium within the electron spin system, followed by equilibration of the electron spin temperature with the crystal lattice temperature.

The results are obtained for Si-doped GaAs samples with uncompensated donor concentrations $n_\text{D} = 5.5 \times 10^{14}$~cm$^{-3}$ (2-$\mu$m-thick layer grown by the molecular-beam epitaxy), $1.0 \times 10^{15}$, $4.0 \times 10^{15}$~cm$^{-3}$ and $1.6 \times 10^{16}$~cm$^{-3}$  (20, 20 and 7-$\mu$m-thick, respectively, layers grown by liquid-phase epitaxy), $1.4 \times 10^{16}$, $3.7 \times 10^{16}$ and $7.1 \times 10^{16}$~cm$^{-3}$ (350, 170 and 170-$\mu$m-thick, respectively, bulk wafers). The samples are placed in the variable temperature insert of a split-coil magnetocryostat ($T =2-25$~K). Magnetic fields up to 6~T are applied either parallel (Faraday geometry) or perpendicular (Voigt geometry) to the light propagation vector that is parallel to the sample normal.

The extended pump-probe Kerr/Faraday rotation technique as described in Ref.~\cite{Belykh2016} is used to study the electron spin dynamics. It is a modification of the standard pump-probe Kerr/Faraday rotation technique, where circularly-polarized pump pulses generate carrier spin polarization, which is then probed by the Kerr(Faraday) rotation of linearly-polarized probe pulses after reflection(transmission) from(through) the sample. Implementation of pulse picking for both pump and probe beams in combination with a mechanical delay line allows us to scan microsecond time ranges with picosecond time resolution. Details of the technique are given in Ref.~\cite{Belykh2016}.
Here, a Ti:Sapphire laser emits a train of 2~ps pulses with a repetition rate of 76~MHz (repetition period $T_\text{R}=13.1$~ns). The pump protocol uses single pulses per excitation period. The separation between these pulses is $80 T_\text{R}$, $160 T_\text{R}$ or $320 T_\text{R}$ in order to clearly exceed the characteristic time of spin polarization decay. The samples with donor concentrations $n_\text{D}$ of $5.5 \times 10^{14}$, $1.0 \times 10^{15}$, $4.0 \times 10^{15}$~cm$^{-3}$ and $1.6 \times 10^{16}$~cm$^{-3}$ are studied in reflection geometry (Kerr rotation) with the laser wavelength set to 819~nm, close to the donor-bound exciton resonance. The samples with $n_\text{D} = 1.4 \times 10^{16}$, $3.7 \times 10^{16}$ and $7.1 \times 10^{16}$~cm$^{-3}$ are studied in transmission geometry (Faraday rotation) with the laser wavelength set to 825, 829 and 829~nm, respectively.

\begin{figure*}
\includegraphics[width=2\columnwidth]{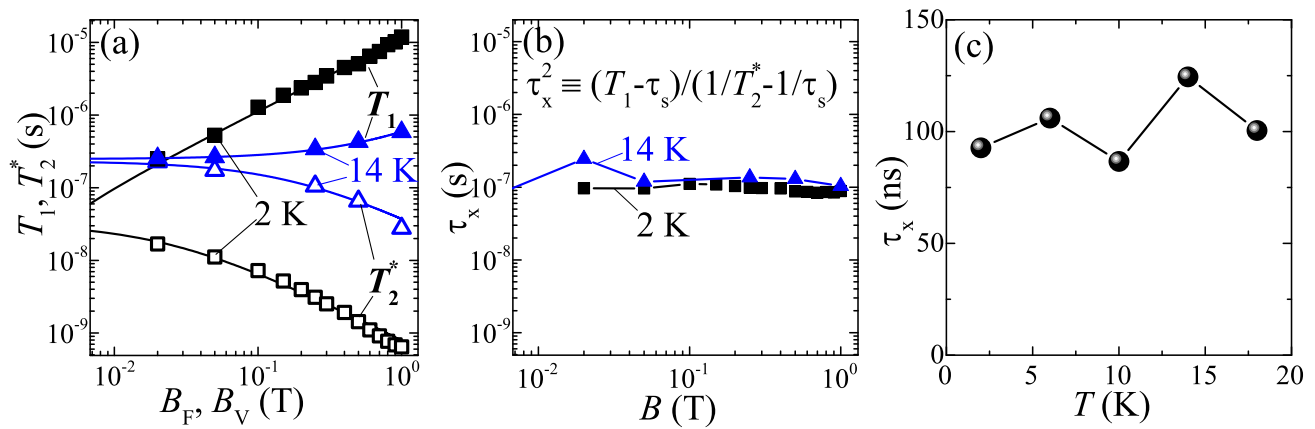}
\caption{(a) Magnetic field dependence of $T_1$ (solid symbols) and $T_2^*$ (open symbols) at different temperatures.  Lines show linear fits to $T_1$ and to the reciprocal of $T_2^*$. (b) Magnetic field dependence of time $\tau_\text{x}$ defined according to the Eq.~\eqref{eq:taux} for different temperatures. (c) Temperature dependence of $\tau_\text{x}$. (a)-(c) $n_\text{D} = 5.5 \times 10^{14}$~cm$^{-3}$.}
\label{fig:taux}
\end{figure*}

First we study the effect of electron delocalization on the inhomogeneous dephasing of the spin ensemble. Figure~\ref{FigVoigt}(a) shows the dynamics of the electron spin precession about different magnetic fields $B_\text{V}$ applied in the Voigt geometry for the weakly doped sample ($n_\text{D} = 5.5 \times 10^{14}$~cm$^{-3}$) at $T = 2$~K, where almost all electrons are localized on donors. In weak magnetic fields the spin precession decays with the time $T_2^* \approx 30$~ns, in good agreement with Hanle-effect measurements \cite{Dzhioev2002}. This inhomogeneous decay is determined by the ensemble-averaged electron spin precession about the random nuclear fields in the vicinity of donors. With increasing $B_\text{V}$ the dynamics becomes considerably shorter, so that the spin dephasing time $T_2^*$ rapidly decreases to $\sim 1$~ns at $B_\text{V} = 500$~mT [see Fig.~\ref{fig:taux}(a), open squares]. This decrease is well described with the equation $1/T_2^* = 1/\tau_\text{s} + \delta g \mu_\text{B} B_\text{V} / \hbar$ [$\tau_\text{s} = T_2^*(B=0)$ and $\mu_\text{B}$ is the Bohr magneton], indicating a large spread of electron $g$ factors $\delta g \approx 1.4 \times 10^{-2}$. This spread arises from the $g$ factor variation of electrons bound by donors that are located at different positions. One can estimate the variation from the spread of localized electron energies, $\delta E \sim 1$~meV using the Roth-Lax-Zwerdling equation \cite{Roth1959}, which gives a $\delta g \sim 10^{-2}$ in agreement with the experiment.

It is straightforward to delocalize electrons in this sample by increasing the lattice temperature. Surprisingly, at $T = 14$~K [Fig.~\ref{FigVoigt}(b)] the dynamics shows a much slower decay with $T_2^* \approx 220$~ns at weak magnetic fields. With increasing $B_\text{V}$ the dynamics continues to stay considerably longer than at $T = 2$~K. It also shows slow beatings with a frequency linearly increasing with magnetic field, indicating a $g$-factor splitting of $\Delta g \approx 1.5 \times 10^{-2}$. This splitting may be related to different electron subensembles, e.g. localized and free electrons. The magnetic field dependence of $T_2^*$ for the dominating component in the beating signal [Fig.~\ref{fig:taux}(a), open triangles] gives $\delta g = 2 \times 10^{-4}$, drastically smaller than at $T = 2$~K. The temperature dependence of the $g$-factor spread for this lightly doped sample is shown in Fig.~\ref{FigVoigt}(d). As $T$ is increased from 2 to 18~K, $\delta g$ monotonically decreases by almost two orders of magnitude.

The strong suppression of the inhomogeneous spin dephasing as result of electron delocalization is also found for the sample with donor concentration $n_\text{D} = 1.4 \times 10^{16}$~cm$^{-3}$, close to the MIT, where a considerable fraction of the electrons is already delocalized at low $T$. The electron spin precession dynamics of this sample [Fig.~\ref{FigVoigt}(c)] shows a long dephasing time at weak fields, $T_2^*(B = 0) \approx 250$~ns, and a small $\delta g \sim 2 \times 10^{-4}$ even at low temperatures \cite{Belykh2016}.

\begin{figure}
\includegraphics[width=1\columnwidth]{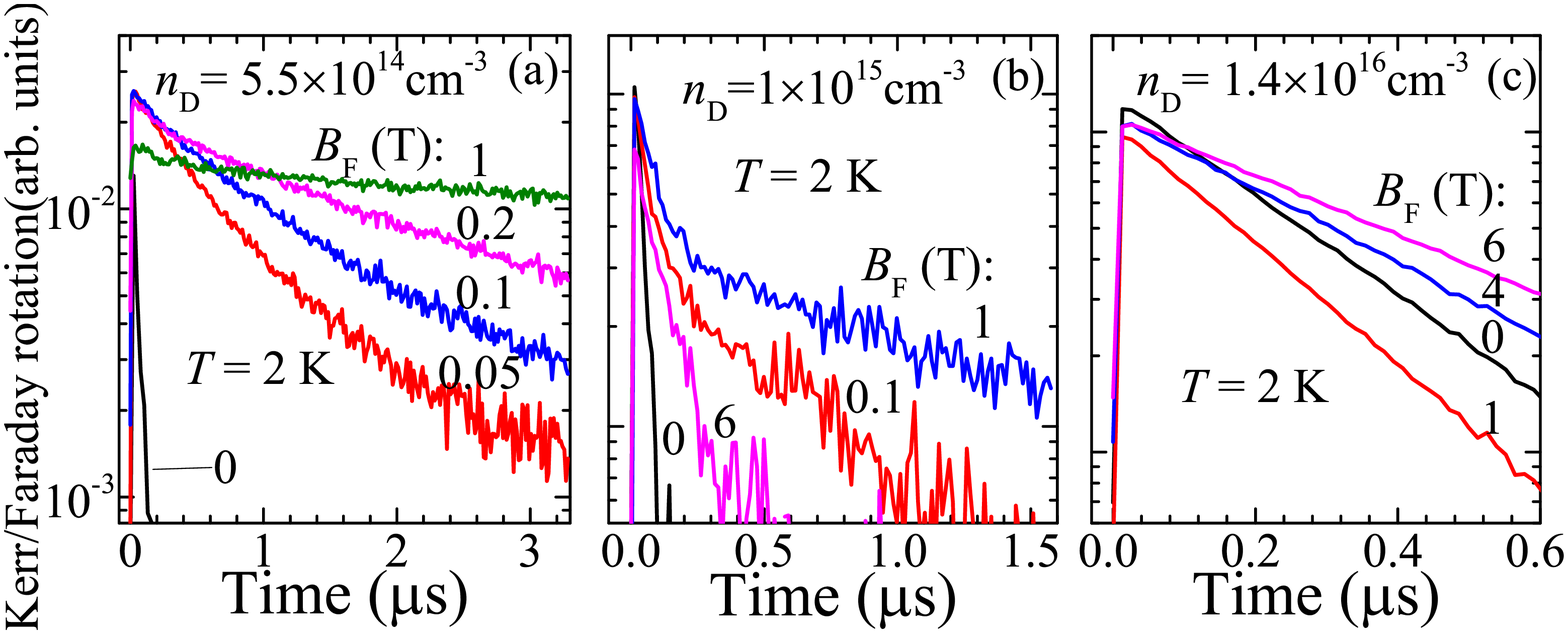}
\caption{(a),(b),(c) Dynamics of Kerr/Faraday rotation signal for different magnetic fields applied in Faraday geometry for samples with different donor concentrations. $T = 2$~K.}
\label{FigFaraday}
\end{figure}

In order to study the longitudinal spin dynamics, we apply the magnetic field $B_\text{F}$ perpendicular to the sample surface and parallel to the optical axis (Faraday geometry). Figure~\ref{FigFaraday}(a) shows the longitudinal spin relaxation dynamics for the weakly doped sample ($n_\text{D} = 5.5 \times 10^{14}$~cm$^{-3}$) for different magnetic fields $B_\text{F}$ at $T = 2$~K. For $B_\text{F} \approx 0$ the dynamics shows a decay with $T_1 \approx 30$~ns in good agreement with the $T_2^*$ value obtained above in a weak Voigt field. Note that at $B = 0$, $T_1 = T_2^*$. With increasing magnetic field the dynamics becomes much slower and can be described by a double-exponential decay with a weak fast component with a decay time of $\sim 300$~ns and a dominating slow component with a $T_1$ time that strongly depends on $B_\text{F}$ [see Fig.~\ref{fig:taux}(a), solid squares]. $T_1$ increases with $B_\text{F}$ from 30~ns to 12~$\mu$s. This increase is almost linear across the studied range of $B_\text{F}$ from 0 to 1~T. Above $B_\text{F} = 1$~T the signal of this sample becomes hardly detectable.

For the sample with a higher donor concentration of $1 \times 10^{15}$~cm$^{-3}$, however, still below the MIT, the longitudinal spin dynamics clearly shows a double-exponential decay with a much more pronounced fast component [Fig.~\ref{FigFaraday}(b)]. For the sample with $n_\text{D} = 1.4 \times 10^{16}$~cm$^{-3}$, just around the MIT, the spin dynamics is single-exponential for all studied magnetic fields [Fig.~\ref{FigFaraday}(c)]. The magnetic field dependence of $T_1=200-500$~ns is much weaker than that for the low donor concentration samples.

\begin{figure*}[t]
\includegraphics[width=1.5\columnwidth]{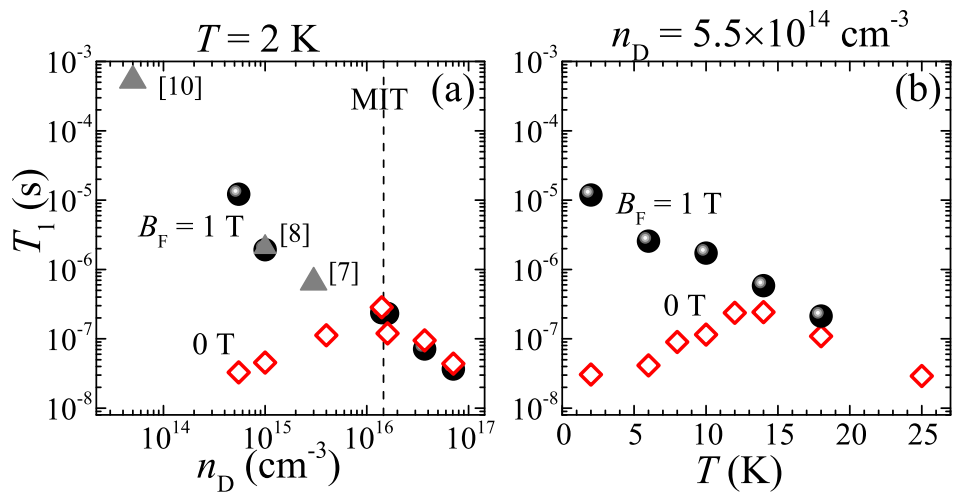}
\caption{(a) Dependence of the longitudinal spin relaxation time $T_1$ on donor concentration for $B_\text{F} =$0~T (open symbols) and 1~T (solid symbols). Triangles correspond to literature data \cite{Colton2004,Colton2007,Linpeng2016}. $T=2$~K. (b) Dependence of $T_1$ on lattice temperature for $B_\text{F} =$ 0~T (open symbols) and 1~T (solid symbols) in the sample with $n_\text{D} = 5.5 \times 10^{14}$~cm$^{-3}$.}
\label{FignTDep}
\end{figure*}

Figure~\ref{FignTDep} summarizes the longitudinal relaxation times at $B_\text{F}=0$ and 1~T as function of donor concentration [Fig.~\ref{FignTDep}(a)] and temperature [Fig.~\ref{FignTDep}(b)]. Figure~\ref{FignTDep}(a) also includes literature data for $T_1$ at 1~T (the triangles) measured by pump-probe methods analysing the polarization-resolved photoluminescence \cite{Colton2004,Colton2007,Linpeng2016}.
$T_1$ at $B_\text{F} = 1$~T monotonically decreases with $n_\text{D}$ by three orders of magnitude without a change of this trend at the MIT threshold.
On the other hand, at zero magnetic field $T_1$ first increases and then decreases above  the MIT concentration.
The most striking observation is that the $T_1$ at 0~T and 1~T almost coincide above the MIT, even though they differ distinctly by a few orders of magnitude for low donor concentrations.

The described behavior also applies to the dependence of $T_1$ on temperature [Fig.~\ref{FignTDep}(b)] for the sample with a low donor concentration of $5.5 \times 10^{14}$~cm$^{-3}$, indicating the delocalization onset at 14~K. Above this temperature, $T_1$ is weakly dependent on magnetic field. Interestingly, an increase of the donor concentration from $\sim 5\times 10^{14}$ to $\sim 10^{17}$~cm$^{-3}$ at $T = 2$~K [Fig.~\ref{FignTDep}(a)] has the same effect on the spin relaxation as the increase of temperature from 2 to 25 K for $n_\text{D} = 5.5 \times 10^{14}$~cm$^{-3}$ [Fig.~\ref{FignTDep}(b)], suggesting that $T_1$ shows similar dependences on $\ln(n_\text{D})$ and on $T$.

We found a striking similarity in the enhancement of $T_1$ and the suppression of $T_2$ by a magnetic field as demonstrated in Fig.~\ref{fig:taux}(a). This relation can be described by the quantity 
\begin{equation}
\tau_\text{x}\equiv\sqrt{\frac{T_1-\tau_\text{s}}{1/T_2^*-1/\tau_\text{s}}}\approx \sqrt{T_1 T_2^*},
\label{eq:taux}
\end{equation}
where $\tau_\text{s}=T_1(B=0)=T_2^*(B=0)$.
$\tau_\text{x}$ remains almost constant in the whole studied range of magnetic fields [Fig.\ref{fig:taux}(b)] and temperatures [Fig.\ref{fig:taux}(c)], while $T_1$ and $T_2^*$ change by more than two orders of magnitude.

Let us now discuss the transverse and longitudinal spin relaxation when changing the electron concentration and/or temperature, and how it is related to the localization of electron charge and spin. Note that the electron spin may diffuse even in a system of localized electrons (without charge mobility).

The drastic suppression of the inhomogeneous spin dephasing when electrons become delocalized can be explained by the motional narrowing effect \cite{Pines1955}, assisted by exchange interaction \cite{Paget1981}. When a fraction of electrons is mobile, they transfer spin between donor centers via exchange coupling, reducing the spin correlation time at the individual donors, $\tau_\text{c}$ \cite{Dzhioev2002a}. As a result, the $g$ factor averages over many donors, leading to a narrowing of the total $g$-factor distribution. This is likely to be the main mechanism of the $\delta g$ narrowing with increasing temperature (Fig.~\ref{FigVoigt}). The increase of donor concentration can also result in such narrowing by direct inter-donor exchange, which results in a drastic shortening of $\tau_\text{c}$ and a donor site averaging by spin diffusion \cite{Dzhioev2002,Kavokin2008}. In both cases (and especially in the latter one) it is the \emph{spin} rather than the charge mobility that results in motional narrowing and increase of $T_2^*$. At donor concentrations above the MIT, the electrons are mostly mobile and the $g$-factor broadening due to site inhomogeneity disappears almost completely. In a similar way, delocalization reduces the interaction with nuclear fields that determines the spin dephasing at $B_\text{V} = 0$.
For a further temperature or concentration increase the spin-orbit relaxation becomes dominant for mobile electrons, leading to a decrease of the spin relaxation time (Fig.~\ref{FignTDep}).

The longitudinal spin dynamics of localized electrons in magnetic field is governed by relaxation of both angular momentum and energy \cite{Kavokin2008}. In strong magnetic field, the energy relaxation is of prime importance, since the spin-phonon coupling at low temperatures typically is weak, and the transfer of the Zeeman energy, associated with a spin flip, to the crystal lattice may take a long time. This fact explains the difference of the longitudinal spin relaxation times measured in zero and strong field for all our samples below $n_\text{D}=1.4 \times 10^{16}$~cm$^{-3}$ [Fig.~\ref{FignTDep}(a)]. Since this concentration is close to the MIT, we suggest that at this and higher concentrations the Zeeman energy is efficiently transferred to the motional degrees of freedom of the mobile electrons. Therefore, in contrast to the transverse relaxation, the longitudinal spin relaxation indicates the onset of electron \emph{charge} mobility - as a result of either the MIT with increasing doping or due to thermal activation at elevated $T$.

The unusual inverse relationship between the times $T_1$ and $T_2^*$ as function of magnetic field and temperature [Fig.~\ref{fig:taux}], most pronounced for the lightest-doped sample, is, in fact, a signature of diffusion-limited spin-lattice relaxation that is typical for nuclear spins in solids but has never been found before for electrons. Indeed, neither the hyperfine nor the anisotropic-exchange mechanisms \cite{Dzhioev2002} can provide energy transfer to the lattice. According to \cite{Kavokin2008}, such a transfer may occur through electron hopping within closely spaced (optimal) pairs of charged and neutral donors, which play the role of spin relaxation (killing) centers. Because of the small number of such pairs, the spin-lattice relaxation of the majority of electrons is determined by spin diffusion towards the pairs, mediated by exchange-induced flip-flop transitions. In magnetic field, electrons localized at neighboring donors acquire an energy difference because of the spread of their $g$-factors, quenching the spin diffusion and extending $T_1$. This $g$-factors spread also manifests itself in a decreasing $T_2^*$. The theory described in the Appendix explains the linear increase of $T_1$ with $B$ and exactly reproduces the experimentally observed relation:
\begin{equation}
T_1 T_2^* \approx \frac{n_\text{D}}{4n_\text{p}} \frac{\hbar^2}{\langle J^2\rangle}.
\end{equation}
Thus, $T_1 T_2^*$ is independent on $B$ and $T$, but determined by the mean squared exchange constant $\langle J^2 \rangle$ and the optimal pairs concentration $n_\text{p}$.

If the spin-spin exchange interaction of localized electrons is strong (as in $n$-GaAs at $n_\text{D}$ above $10^{15}$~cm$^{-3}$), the equilibrium within the electron spin system may be established much faster than its thermalization with the crystal lattice. In this case, a biexponential polarization dynamics is observed, as it was theoretically predicted in Ref.~\cite{Kavokin2008}. The faster component describes the internal spin-spin equilibration i.e. establishing a spin temperature. While the much slower component arises from the energy transfer to the lattice. This indeed corresponds to the observations at $n_\text{D}=1 \times 10^{15}$~cm$^{-3}$ [Fig.~\ref{FigFaraday}(b)]. At lower $n_\text{D}=5.5 \times 10^{14}$~cm$^{-3}$ the biexponential decay is much less pronounced [Fig.~\ref{FigFaraday}(a)] due to the weaker exchange coupling between donor-bound electrons.

In conclusion, the method of extended-time-delay Faraday/Kerr pump-probe spectroscopy has allowed us to investigate the longitudinal and transverse electron spin relaxation in $n$-doped GaAs with different donor concentrations at varying magnetic fields and temperatures. We found clear manifestations of both spin and charge delocalization with increasing doping and/or temperature. The electron spin system is shown to experience a crossover from a behavior similar to that of paramagnetic centers in dielectrics, characterized by site inhomogeneity and strongly magnetic-field dependent relaxation times, to a behavior in the motional-narrowing regime with a single spin lifetime. Further, unexpectedly the relation $T_1 T_2^*\approx$const was extracted from the data. It shows that while spin ensemble inhomogeneity suppresses $T_2^*$, it surprisingly enhances $T_1$. This relation was theoretically consolidateded by considering the diffusion-limited longitudinal spin relaxation. A range of donor concentrations has been found in which an electron spin temperature, different from the lattice temperature, can be established.

\begin{acknowledgments}
We are grateful to M.~M.~Glazov, A.~Greilich and V.~L.~Korenev for valuable advices and useful discussions and to E.~Evers for technical support. We acknowledge the financial support of the Deutsche Forschungsgemeinschaft in the frame of the ICRC TRR 160 (project A1) and the Russian Foundation for Basic Research (Project No. 15-52-12020). KVK acknowledges Saint-Petersburg State University for a research grant 11.34.2.2012.
\end{acknowledgments}

\section{Appendix: Theory of longitudinal spin relaxation in insulating phase}
\setcounter{equation}{0}
\renewcommand{\theequation}{A\arabic{equation}}

The longitudinal spin relaxation in the insulating phase requires the transfer of the Zeeman energy of donor-bound electrons to the crystal lattice. Neither the hyperfine coupling nor the anisotropic exchange interaction, responsible for the spin relaxation in the impurity band of $n$-GaAs at zero magnetic field \cite{Dzhioev2002}, can provide such a transfer. As discussed in \cite{Kavokin2008}, energy relaxation in the impurity band can occur via optimally spaced (up to approximately 4 donor Bohr radii) pairs of charged and neutral donors. Phonon-assisted electron hopping within such a pair effectively couples the spin system to the phonon bath through modulated exchange interactions with donor-bound electrons nearby. Similarly to the nuclear spin-lattice relaxation via paramagnetic centers, relaxation of electrons remote from optimal pairs should be assisted by spin diffusion. The resulting diffusion-limited relaxation is characterized by the time $T_\text{D}$, given by the relation \cite{DeGennes1958}:
\begin{equation}
T_\text{D}^{-1} = 4\pi D_s a_\text{c} n_\text{c},
\label{eq:TD}
\end{equation}
where $D_s$ is the coefficient of spin diffusion, $a_\text{c}$ is the effective interaction radius of the center, and $n_\text{c}$ is the concentration of centers. To apply Eq.~\eqref{eq:TD} to the case of relaxation by optimal pairs, one should substitute the concentration of such pairs $n_\text{p}$ for the number of centers, as well as the average inter-donor distance $n_\text{D}^{-1/3}$ for the interaction radius, and use the expression for the coefficient of the exchange-mediated spin diffusion \cite{Kavokin2008}:
\begin{equation}
D_s = \frac{1}{3} n_\text{D}^{-2/3} \tau_\text{c}^{-1},
\label{eq:Ds}
\end{equation}
where $\tau_\text{c}$ is the spin correlation time at a donor, determined by the isotropic exchange interaction with other donor-bound electrons. In doing so, we obtain:
\begin{equation}
T_1 \approx T_\text{D} \approx  \frac{n_\text{D}}{4n_\text{p}} \tau_\text{c},
\label{eq:TDfin}
\end{equation}
(the same relation save for the insignificant factor $1/4$ was obtained in \cite{Kavokin2008}).
The correlation time $\tau_\text{c}$ is determined by the spectral power density $b_\text{ex}^2(\omega)$ of fluctuating fields acting upon an electron spin due to its exchange interaction with its neighbors \cite{Abragam1961}. In a longitudinal field, when $T_2^* \ll T_2$, it is determined by the spread of Larmor frequencies of the electron spins:
\begin{multline}
\frac{1}{\tau_\text{c}}\approx\frac{(\langle g\rangle\mu_\text{B})^2}{\hbar^2}\langle b_\text{ex}^2(\omega)\rangle_g=\\
\frac{\langle J^2 \rangle}{\hbar^2} \frac{\hbar}{\mu_\text{B} B}
\int_{-\infty}^{\infty}\rho_g^2(\frac{\hbar\omega}{\mu_\text{B}B})dg=\\
\frac{\langle J^2 \rangle}{\hbar^2} \frac{\hbar}{\delta g \mu_\text{B} B}= \frac{\hbar}{\tau_\text{c0}^2 \delta g \mu_\text{B} B},
\label{eq:long}
\end{multline}
where $\rho_g$ is the distribution function of the electron $g$-factor, $\delta g = (\int_{-\infty}^{\infty}\rho_g^2(g)dg)^{-1}$, $\langle J^2 \rangle$  is the mean squared exchange constant and $\tau_{c0} = \hbar /\sqrt{\langle J^2 \rangle}$ is the correlation time at zero magnetic field and temperature. Finally we obtain:
\begin{equation}
T_1 \approx \frac{n_\text{D}}{4n_\text{p}} \tau_\text{c0}^2 \frac{\delta g \mu_\text{B} B}{\hbar}.
\end{equation}
Thus, $T_1 \propto B$ in agreement with the experiment. Taking into account
\begin{equation}
1/T_2^* = 1/\tau_\text{s} + \delta g \mu_\text{B} B / \hbar \approx \delta g \mu_\text{B} B / \hbar,
\label{EqdivT2}
\end{equation}
which also well describes the experimental dependence in transverse magnetic field, we obtain
\begin{equation}
\tau_\text{x} \approx \sqrt{T_1 T_2^*}\approx \sqrt{\frac{n_\text{D}}{4n_\text{p}}} \tau_\text{c0}.
\end{equation}
The exchange-mediated correlation time $\tau_\text{c0}$ was calculated in Ref.~\cite{Dzhioev2002}. For $n_\text{D} = 5.5 \times 10^{14}$~cm$^{-3}$, $\tau_\text{c0}\approx 10$~ns. To achieve the experimental value $\tau_\text{x} \approx 100$~ns, one should assume a concentration of optimal pairs $n_\text{p} \approx 2.5 \times 10^{-3} n_\text{D}$ in a reasonable agreement with theoretical estimate in Ref.~\cite{Kavokin2008}.

The presented model suggests that the link between $T_1$ and $T_2^*$ is determined by an exchange interaction strength (via the spin correlation time) which is independent on magnetic field and temperature. Both $T_1$ and $T_2^*$ are governed by the spin ensemble inhomogeneity. The larger the inhomogeneity is, the smaller is $T_2^*$ and the larger is $T_1$.


\begin{thebibliography}{18}%
\makeatletter
\providecommand \@ifxundefined [1]{%
 \@ifx{#1\undefined}
}%
\providecommand \@ifnum [1]{%
 \ifnum #1\expandafter \@firstoftwo
 \else \expandafter \@secondoftwo
 \fi
}%
\providecommand \@ifx [1]{%
 \ifx #1\expandafter \@firstoftwo
 \else \expandafter \@secondoftwo
 \fi
}%
\providecommand \natexlab [1]{#1}%
\providecommand \enquote  [1]{``#1''}%
\providecommand \bibnamefont  [1]{#1}%
\providecommand \bibfnamefont [1]{#1}%
\providecommand \citenamefont [1]{#1}%
\providecommand \href@noop [0]{\@secondoftwo}%
\providecommand \href [0]{\begingroup \@sanitize@url \@href}%
\providecommand \@href[1]{\@@startlink{#1}\@@href}%
\providecommand \@@href[1]{\endgroup#1\@@endlink}%
\providecommand \@sanitize@url [0]{\catcode `\\12\catcode `\$12\catcode
  `\&12\catcode `\#12\catcode `\^12\catcode `\_12\catcode `\%12\relax}%
\providecommand \@@startlink[1]{}%
\providecommand \@@endlink[0]{}%
\providecommand \url  [0]{\begingroup\@sanitize@url \@url }%
\providecommand \@url [1]{\endgroup\@href {#1}{\urlprefix }}%
\providecommand \urlprefix  [0]{URL }%
\providecommand \Eprint [0]{\href }%
\providecommand \doibase [0]{http://dx.doi.org/}%
\providecommand \selectlanguage [0]{\@gobble}%
\providecommand \bibinfo  [0]{\@secondoftwo}%
\providecommand \bibfield  [0]{\@secondoftwo}%
\providecommand \translation [1]{[#1]}%
\providecommand \BibitemOpen [0]{}%
\providecommand \bibitemStop [0]{}%
\providecommand \bibitemNoStop [0]{.\EOS\space}%
\providecommand \EOS [0]{\spacefactor3000\relax}%
\providecommand \BibitemShut  [1]{\csname bibitem#1\endcsname}%
\let\auto@bib@innerbib\@empty
\bibitem [{\citenamefont {Dzhioev}\ \emph
  {et~al.}(2002{\natexlab{a}})\citenamefont {Dzhioev}, \citenamefont {Kavokin},
  \citenamefont {Korenev}, \citenamefont {Lazarev}, \citenamefont {Meltser},
  \citenamefont {Stepanova}, \citenamefont {Zakharchenya}, \citenamefont
  {Gammon},\ and\ \citenamefont {Katzer}}]{Dzhioev2002}%
  \BibitemOpen
  \bibfield  {author} {\bibinfo {author} {\bibfnamefont {R.~I.}\ \bibnamefont
  {Dzhioev}}, \bibinfo {author} {\bibfnamefont {K.~V.}\ \bibnamefont
  {Kavokin}}, \bibinfo {author} {\bibfnamefont {V.~L.}\ \bibnamefont
  {Korenev}}, \bibinfo {author} {\bibfnamefont {M.~V.}\ \bibnamefont
  {Lazarev}}, \bibinfo {author} {\bibfnamefont {B.~Y.}\ \bibnamefont
  {Meltser}}, \bibinfo {author} {\bibfnamefont {M.~N.}\ \bibnamefont
  {Stepanova}}, \bibinfo {author} {\bibfnamefont {B.~P.}\ \bibnamefont
  {Zakharchenya}}, \bibinfo {author} {\bibfnamefont {D.}~\bibnamefont
  {Gammon}}, \ and\ \bibinfo {author} {\bibfnamefont {D.~S.}\ \bibnamefont
  {Katzer}},\ }\href {\doibase 10.1103/PhysRevB.66.245204} {\bibfield
  {journal} {\bibinfo  {journal} {Phys. Rev. B}\ }\textbf {\bibinfo {volume}
  {66}},\ \bibinfo {pages} {245204} (\bibinfo {year}
  {2002}{\natexlab{a}})}\BibitemShut {NoStop}%
\bibitem [{\citenamefont {Crooker}\ \emph {et~al.}(2009)\citenamefont
  {Crooker}, \citenamefont {Cheng},\ and\ \citenamefont {Smith}}]{Crooker2009}%
  \BibitemOpen
  \bibfield  {author} {\bibinfo {author} {\bibfnamefont {S.~A.}\ \bibnamefont
  {Crooker}}, \bibinfo {author} {\bibfnamefont {L.}~\bibnamefont {Cheng}}, \
  and\ \bibinfo {author} {\bibfnamefont {D.~L.}\ \bibnamefont {Smith}},\ }\href
  {\doibase 10.1103/PhysRevB.79.035208} {\bibfield  {journal} {\bibinfo
  {journal} {Phys. Rev. B}\ }\textbf {\bibinfo {volume} {79}},\ \bibinfo
  {pages} {035208} (\bibinfo {year} {2009})}\BibitemShut {NoStop}%
\bibitem [{\citenamefont {R\"{o}mer}\ \emph {et~al.}(2010)\citenamefont
  {R\"{o}mer}, \citenamefont {Bernien}, \citenamefont {M\"{u}ller},
  \citenamefont {Schuh}, \citenamefont {H\"{u}bner},\ and\ \citenamefont
  {Oestreich}}]{Romer2010}%
  \BibitemOpen
  \bibfield  {author} {\bibinfo {author} {\bibfnamefont {M.}~\bibnamefont
  {R\"{o}mer}}, \bibinfo {author} {\bibfnamefont {H.}~\bibnamefont {Bernien}},
  \bibinfo {author} {\bibfnamefont {G.}~\bibnamefont {M\"{u}ller}}, \bibinfo
  {author} {\bibfnamefont {D.}~\bibnamefont {Schuh}}, \bibinfo {author}
  {\bibfnamefont {J.}~\bibnamefont {H\"{u}bner}}, \ and\ \bibinfo {author}
  {\bibfnamefont {M.}~\bibnamefont {Oestreich}},\ }\href {\doibase
  10.1103/PhysRevB.81.075216} {\bibfield  {journal} {\bibinfo  {journal} {Phys.
  Rev. B}\ }\textbf {\bibinfo {volume} {81}},\ \bibinfo {pages} {075216}
  (\bibinfo {year} {2010})}\BibitemShut {NoStop}%
\bibitem [{\citenamefont {Furis}\ \emph {et~al.}(2006)\citenamefont {Furis},
  \citenamefont {Smith}, \citenamefont {Crooker},\ and\ \citenamefont
  {Reno}}]{Furis2006}%
  \BibitemOpen
  \bibfield  {author} {\bibinfo {author} {\bibfnamefont {M.}~\bibnamefont
  {Furis}}, \bibinfo {author} {\bibfnamefont {D.~L.}\ \bibnamefont {Smith}},
  \bibinfo {author} {\bibfnamefont {S.~A.}\ \bibnamefont {Crooker}}, \ and\
  \bibinfo {author} {\bibfnamefont {J.~L.}\ \bibnamefont {Reno}},\ }\href
  {\doibase 10.1063/1.2345608} {\bibfield  {journal} {\bibinfo  {journal}
  {Appl. Phys. Lett.}\ }\textbf {\bibinfo {volume} {89}},\ \bibinfo {pages}
  {102102} (\bibinfo {year} {2006})}\BibitemShut {NoStop}%
\bibitem [{\citenamefont {Lonnemann}\ \emph {et~al.}(2017)\citenamefont
  {Lonnemann}, \citenamefont {Rugeramigabo}, \citenamefont {Oestreich},\ and\
  \citenamefont {H\"{u}bner}}]{Lonnemann2017}%
  \BibitemOpen
  \bibfield  {author} {\bibinfo {author} {\bibfnamefont {J.~G.}\ \bibnamefont
  {Lonnemann}}, \bibinfo {author} {\bibfnamefont {E.~P.}\ \bibnamefont
  {Rugeramigabo}}, \bibinfo {author} {\bibfnamefont {M.}~\bibnamefont
  {Oestreich}}, \ and\ \bibinfo {author} {\bibfnamefont {J.}~\bibnamefont
  {H\"{u}bner}},\ }\href {\doibase 10.1103/PhysRevB.96.045201} {\bibfield
  {journal} {\bibinfo  {journal} {Phys. Rev. B}\ }\textbf {\bibinfo {volume}
  {96}},\ \bibinfo {pages} {045201} (\bibinfo {year} {2017})}\BibitemShut
  {NoStop}%
\bibitem [{\citenamefont {Kikkawa}\ and\ \citenamefont
  {Awschalom}(1998)}]{Kikkawa1998}%
  \BibitemOpen
  \bibfield  {author} {\bibinfo {author} {\bibfnamefont {J.~M.}\ \bibnamefont
  {Kikkawa}}\ and\ \bibinfo {author} {\bibfnamefont {D.~D.}\ \bibnamefont
  {Awschalom}},\ }\href {\doibase 10.1103/PhysRevLett.80.4313} {\bibfield
  {journal} {\bibinfo  {journal} {Phys. Rev. Lett.}\ }\textbf {\bibinfo
  {volume} {80}},\ \bibinfo {pages} {4313} (\bibinfo {year}
  {1998})}\BibitemShut {NoStop}%
\bibitem [{\citenamefont {Colton}\ \emph {et~al.}(2004)\citenamefont {Colton},
  \citenamefont {Kennedy}, \citenamefont {Bracker},\ and\ \citenamefont
  {Gammon}}]{Colton2004}%
  \BibitemOpen
  \bibfield  {author} {\bibinfo {author} {\bibfnamefont {J.~S.}\ \bibnamefont
  {Colton}}, \bibinfo {author} {\bibfnamefont {T.~A.}\ \bibnamefont {Kennedy}},
  \bibinfo {author} {\bibfnamefont {A.~S.}\ \bibnamefont {Bracker}}, \ and\
  \bibinfo {author} {\bibfnamefont {D.}~\bibnamefont {Gammon}},\ }\href
  {\doibase 10.1103/PhysRevB.69.121307} {\bibfield  {journal} {\bibinfo
  {journal} {Phys. Rev. B}\ }\textbf {\bibinfo {volume} {69}},\ \bibinfo
  {pages} {121307} (\bibinfo {year} {2004})}\BibitemShut {NoStop}%
\bibitem [{\citenamefont {Colton}\ \emph {et~al.}(2007)\citenamefont {Colton},
  \citenamefont {Heeb}, \citenamefont {Schroeder}, \citenamefont {Stokes},
  \citenamefont {Wienkes},\ and\ \citenamefont {Bracker}}]{Colton2007}%
  \BibitemOpen
  \bibfield  {author} {\bibinfo {author} {\bibfnamefont {J.~S.}\ \bibnamefont
  {Colton}}, \bibinfo {author} {\bibfnamefont {M.~E.}\ \bibnamefont {Heeb}},
  \bibinfo {author} {\bibfnamefont {P.}~\bibnamefont {Schroeder}}, \bibinfo
  {author} {\bibfnamefont {A.}~\bibnamefont {Stokes}}, \bibinfo {author}
  {\bibfnamefont {L.~R.}\ \bibnamefont {Wienkes}}, \ and\ \bibinfo {author}
  {\bibfnamefont {A.~S.}\ \bibnamefont {Bracker}},\ }\href {\doibase
  10.1103/PhysRevB.75.205201} {\bibfield  {journal} {\bibinfo  {journal} {Phys.
  Rev. B}\ }\textbf {\bibinfo {volume} {75}},\ \bibinfo {pages} {205201}
  (\bibinfo {year} {2007})}\BibitemShut {NoStop}%
\bibitem [{\citenamefont {Fu}\ \emph {et~al.}(2006)\citenamefont {Fu},
  \citenamefont {Yeo}, \citenamefont {Clark}, \citenamefont {Santori},
  \citenamefont {Stanley}, \citenamefont {Holland},\ and\ \citenamefont
  {Yamamoto}}]{Fu2006}%
  \BibitemOpen
  \bibfield  {author} {\bibinfo {author} {\bibfnamefont {K.-M.~C.}\
  \bibnamefont {Fu}}, \bibinfo {author} {\bibfnamefont {W.}~\bibnamefont
  {Yeo}}, \bibinfo {author} {\bibfnamefont {S.}~\bibnamefont {Clark}}, \bibinfo
  {author} {\bibfnamefont {C.}~\bibnamefont {Santori}}, \bibinfo {author}
  {\bibfnamefont {C.}~\bibnamefont {Stanley}}, \bibinfo {author} {\bibfnamefont
  {M.~C.}\ \bibnamefont {Holland}}, \ and\ \bibinfo {author} {\bibfnamefont
  {Y.}~\bibnamefont {Yamamoto}},\ }\href {\doibase 10.1103/PhysRevB.74.121304}
  {\bibfield  {journal} {\bibinfo  {journal} {Phys. Rev. B}\ }\textbf {\bibinfo
  {volume} {74}},\ \bibinfo {pages} {121304} (\bibinfo {year}
  {2006})}\BibitemShut {NoStop}%
\bibitem [{\citenamefont {Linpeng}\ \emph {et~al.}(2016)\citenamefont
  {Linpeng}, \citenamefont {Karin}, \citenamefont {Durnev}, \citenamefont
  {Barbour}, \citenamefont {Glazov}, \citenamefont {Sherman}, \citenamefont
  {Watkins}, \citenamefont {Seto},\ and\ \citenamefont {Fu}}]{Linpeng2016}%
  \BibitemOpen
  \bibfield  {author} {\bibinfo {author} {\bibfnamefont {X.}~\bibnamefont
  {Linpeng}}, \bibinfo {author} {\bibfnamefont {T.}~\bibnamefont {Karin}},
  \bibinfo {author} {\bibfnamefont {M.~V.}\ \bibnamefont {Durnev}}, \bibinfo
  {author} {\bibfnamefont {R.}~\bibnamefont {Barbour}}, \bibinfo {author}
  {\bibfnamefont {M.~M.}\ \bibnamefont {Glazov}}, \bibinfo {author}
  {\bibfnamefont {E.~Y.}\ \bibnamefont {Sherman}}, \bibinfo {author}
  {\bibfnamefont {S.~P.}\ \bibnamefont {Watkins}}, \bibinfo {author}
  {\bibfnamefont {S.}~\bibnamefont {Seto}}, \ and\ \bibinfo {author}
  {\bibfnamefont {K.-M.~C.}\ \bibnamefont {Fu}},\ }\href {\doibase
  10.1103/PhysRevB.94.125401} {\bibfield  {journal} {\bibinfo  {journal} {Phys.
  Rev. B}\ }\textbf {\bibinfo {volume} {94}},\ \bibinfo {pages} {125401}
  (\bibinfo {year} {2016})}\BibitemShut {NoStop}%
\bibitem [{\citenamefont {Belykh}\ \emph {et~al.}(2016)\citenamefont {Belykh},
  \citenamefont {Evers}, \citenamefont {Yakovlev}, \citenamefont {Fobbe},
  \citenamefont {Greilich},\ and\ \citenamefont {Bayer}}]{Belykh2016}%
  \BibitemOpen
  \bibfield  {author} {\bibinfo {author} {\bibfnamefont {V.~V.}\ \bibnamefont
  {Belykh}}, \bibinfo {author} {\bibfnamefont {E.}~\bibnamefont {Evers}},
  \bibinfo {author} {\bibfnamefont {D.~R.}\ \bibnamefont {Yakovlev}}, \bibinfo
  {author} {\bibfnamefont {F.}~\bibnamefont {Fobbe}}, \bibinfo {author}
  {\bibfnamefont {A.}~\bibnamefont {Greilich}}, \ and\ \bibinfo {author}
  {\bibfnamefont {M.}~\bibnamefont {Bayer}},\ }\href {\doibase
  10.1103/PhysRevB.94.241202} {\bibfield  {journal} {\bibinfo  {journal} {Phys.
  Rev. B}\ }\textbf {\bibinfo {volume} {94}},\ \bibinfo {pages} {241202}
  (\bibinfo {year} {2016})}\BibitemShut {NoStop}%
\bibitem [{\citenamefont {Dzhioev}\ \emph
  {et~al.}(2002{\natexlab{b}})\citenamefont {Dzhioev}, \citenamefont {Korenev},
  \citenamefont {Merkulov}, \citenamefont {Zakharchenya}, \citenamefont
  {Gammon}, \citenamefont {Efros},\ and\ \citenamefont
  {Katzer}}]{Dzhioev2002a}%
  \BibitemOpen
  \bibfield  {author} {\bibinfo {author} {\bibfnamefont {R.~I.}\ \bibnamefont
  {Dzhioev}}, \bibinfo {author} {\bibfnamefont {V.~L.}\ \bibnamefont
  {Korenev}}, \bibinfo {author} {\bibfnamefont {I.~A.}\ \bibnamefont
  {Merkulov}}, \bibinfo {author} {\bibfnamefont {B.~P.}\ \bibnamefont
  {Zakharchenya}}, \bibinfo {author} {\bibfnamefont {D.}~\bibnamefont
  {Gammon}}, \bibinfo {author} {\bibfnamefont {A.~L.}\ \bibnamefont {Efros}}, \
  and\ \bibinfo {author} {\bibfnamefont {D.~S.}\ \bibnamefont {Katzer}},\
  }\href {\doibase 10.1103/PhysRevLett.88.256801} {\bibfield  {journal}
  {\bibinfo  {journal} {Phys. Rev. Lett.}\ }\textbf {\bibinfo {volume} {88}},\
  \bibinfo {pages} {256801} (\bibinfo {year} {2002}{\natexlab{b}})}\BibitemShut
  {NoStop}%
\bibitem [{\citenamefont {Roth}\ \emph {et~al.}(1959)\citenamefont {Roth},
  \citenamefont {Lax},\ and\ \citenamefont {Zwerdling}}]{Roth1959}%
  \BibitemOpen
  \bibfield  {author} {\bibinfo {author} {\bibfnamefont {L.~M.}\ \bibnamefont
  {Roth}}, \bibinfo {author} {\bibfnamefont {B.}~\bibnamefont {Lax}}, \ and\
  \bibinfo {author} {\bibfnamefont {S.}~\bibnamefont {Zwerdling}},\ }\href
  {\doibase 10.1103/PhysRev.114.90} {\bibfield  {journal} {\bibinfo  {journal}
  {Phys. Rev.}\ }\textbf {\bibinfo {volume} {114}},\ \bibinfo {pages} {90}
  (\bibinfo {year} {1959})}\BibitemShut {NoStop}%
\bibitem [{\citenamefont {Pines}\ and\ \citenamefont
  {Slichter}(1955)}]{Pines1955}%
  \BibitemOpen
  \bibfield  {author} {\bibinfo {author} {\bibfnamefont {D.}~\bibnamefont
  {Pines}}\ and\ \bibinfo {author} {\bibfnamefont {C.~P.}\ \bibnamefont
  {Slichter}},\ }\href {\doibase 10.1103/PhysRev.100.1014} {\bibfield
  {journal} {\bibinfo  {journal} {Phys. Rev.}\ }\textbf {\bibinfo {volume}
  {100}},\ \bibinfo {pages} {1014} (\bibinfo {year} {1955})}\BibitemShut
  {NoStop}%
\bibitem [{\citenamefont {Paget}(1981)}]{Paget1981}%
  \BibitemOpen
  \bibfield  {author} {\bibinfo {author} {\bibfnamefont {D.}~\bibnamefont
  {Paget}},\ }\href {\doibase 10.1103/PhysRevB.24.3776} {\bibfield  {journal}
  {\bibinfo  {journal} {Phys. Rev. B}\ }\textbf {\bibinfo {volume} {24}},\
  \bibinfo {pages} {3776} (\bibinfo {year} {1981})}\BibitemShut {NoStop}%
\bibitem [{\citenamefont {Kavokin}(2008)}]{Kavokin2008}%
  \BibitemOpen
  \bibfield  {author} {\bibinfo {author} {\bibfnamefont {K.~V.}\ \bibnamefont
  {Kavokin}},\ }\href {\doibase 10.1088/0268-1242/23/11/114009} {\bibfield
  {journal} {\bibinfo  {journal} {Semicond. Sci. Technol.}\ }\textbf {\bibinfo
  {volume} {23}},\ \bibinfo {pages} {114009} (\bibinfo {year}
  {2008})}\BibitemShut {NoStop}%
\bibitem [{\citenamefont {de~Gennes}(1958)}]{DeGennes1958}%
  \BibitemOpen
  \bibfield  {author} {\bibinfo {author} {\bibfnamefont {P.-G.}\ \bibnamefont
  {de~Gennes}},\ }\href {\doibase 10.1016/0022-3697(58)90284-1} {\bibfield
  {journal} {\bibinfo  {journal} {J. Phys. Chem. Solids}\ }\textbf {\bibinfo
  {volume} {7}},\ \bibinfo {pages} {345} (\bibinfo {year} {1958})}\BibitemShut
  {NoStop}%
\bibitem [{\citenamefont {Abragam}(1961)}]{Abragam1961}%
  \BibitemOpen
  \bibfield  {author} {\bibinfo {author} {\bibfnamefont {A.}~\bibnamefont
  {Abragam}},\ }\href@noop {} {\emph {\bibinfo {title} {{The Principles of
  Nuclear Magnetism}}}}\ (\bibinfo  {publisher} {Oxford University Press},\
  \bibinfo {address} {New York},\ \bibinfo {year} {1961})\ pp.\ \bibinfo
  {pages} {272--274}\BibitemShut {NoStop}%
\end{thebibliography}
\end{document}